\documentstyle[12pt]{article}
\begin{document}

\title{ON THE JASTROW WAVE FUNCTION FOR N-PARTICLES ON A PLANE}

\author{A. AZHARI $^{\star,1}$ and T.T. TRUONG $^{\dagger,2}$ \\                                    
 $ \star $) Laboratoire de Mod\`eles de Physique Math\'ematique,\\
      D\'epartement de physique, Universit\'e de Tours,\\
      Parc de Grandmont, F-37200 Tours, France. \\ \\
 $ \dagger $) Groupe de Physique Statistique.\\
         Universit\'e de Cergy-Pontoise,\\                                       F-95302 Cergy-Pontoise Cedex France. }

\maketitle

\begin{abstract}
The possibility of describing  quantum mechanical particles on a 
plane by a Jastrow wave function is studied. We obtain the condition
 under which the particles interact through pair potentials. It is found
 that if the interparticle forces depend only on distance, then only the
 attractive harmonic pair potential is consistent with the Jastrow
 ansatz. We discuss some connections with wave functions encountered
 in the quantized Hall effect. 
\end{abstract}
\noindent $^1${e-mail: azhari@celfi.phys.univ-tours.fr}\\
\noindent $^2${e-mail: truong@u-cergy.fr}

\newpage

1- If traditional many body problems cover areas such as nuclear or atomic 
physics, contemporary ones are concerned with two dimensional condensed
 matter phenomena such as quantum Hall effect, high-$T_{C}$ 
superconductivity, quantum dots in semi-conductors, etc .....$^{1)}$. The
 physical problem is the description of a large number of identical 
particles (e.g. fermions) in the presence of mutual interactions and
 external forces, in their stable macros-\\copic ground state.
 Excited states may
 then be constructed out of those. A frequent ansatz for such ground 
state is the Jastrow wave function as  used in the theory of quantum
 fluids$^{2)}$ which appears as a product of pair wave functions.\\

The Jastrow wave function is generally used as variational wavefunction.
 But in the last 30 years it has regained considerable
 interest in one dimensional physics since it has been shown that such
 type of wave function may describe the exact ground state for a
 large class of particles  interacting through pair potentials. One may
 speak of completely integrable systems generalising the known
 $ \delta $-function pair potentials in one dimension$^{3)}$.\\

In this situation, one may wonder whether the Jastrow wave function
 could have an equal success in two space dimensions. Of course the
 N-body problem in the plane is more complicated and  only partial results
are known $^{4)}$ as well as a recent ``reduction'' to a three body
 problem$^{1)}$. In the follow-\\ing we seek to establish, as in one
 dimension, the condition under which a Jastrow wave function,
 for a system of non-relativistic particles interacting, solely through
 pair potentials. The question of statistics will be discussed later on,
 since it is known that a two dimensional system may admit ``exotic''
 statistics, which are eventually relevant for the quantum Hall
 effect$^{5)}$.\\

2- For a better understanding, let us review the situation in
 one\\ dimension, following F. Calogero$^{6)}$ and B. Sutherland$^{7)}$.
 Starting from the Jastrow wave function for N particles, on the line
 with abscissas $ x_1<x_2<\ldots<x_N $, i.e.
\begin{equation}
\Psi=\displaystyle\prod_{\scriptstyle i<j}|\psi(x_i-x_j)|^\lambda
\end{equation} 
where  $\psi(x)$ is a chosen pair wave function and $ \lambda$ a
 real number,
 one may seek to determine  the net potential under which the N
 particles are moving. Assuming that one deals with particles of unit
 mass, this N-body potential $ V(x_1, x_2, \ldots, x_N)$ appears as a
 sum of 
pair potentials and a sum of 3-body potentials as follows:
\begin{equation}
V(x_1, x_2, \ldots, x_N)=\displaystyle\sum_{\scriptstyle i<j}V_2(x_i-x_j)+\displaystyle
\sum_{\scriptstyle (i,j,k)}V_3(x_i-x_j, x_j-x_k, x_k-x_i)
\end{equation}
 
The potentials $V_2$ and $V_3$ are functions of the pair wave function 
$\psi(x)$ and expressed in terms of its logarithmic derivative
 $\varphi(x)$ as 
\begin{equation}
V_2(x)= \lambda[\varphi'(x)+\lambda\varphi^2(x)]
\end{equation}
and 
\begin{equation}
V_3(x, y, z)= -\lambda^2[\varphi(x)\varphi(y)+\varphi(y)\varphi(z)+
\varphi(z)\varphi(x)]
\end{equation}
with\hskip4truecm $ x+y+z=0 $

Evidently, if $V_2$ appears naturel from the structure of $ \Psi, V_3$ 
is  an ``induced'' 3-body potential among the particles. B. Sutherland
 cleverly observed
that the N-particles will experience only an effective pair potential
 if $V_3$ splits up into a sum of the following type
 \begin{equation}
\varphi(x)\varphi(y)+\varphi(y)\varphi(z)+\varphi(z)\varphi(x)=
 f(x)+f(y)+f(z)
\end{equation}
for  $x+y+z=0$, and where $f(x)$ is a function obtained from special\\ 
properties of $\psi $. \\
          
 The effective pair potential is then of the form:
\begin{equation}
 V_{2}^{*}(x)=V_2(x)-\lambda^2f(x)
\end{equation}

The condition (5) found by B. Sutherland originally$^{7)}$ is verified by
 a wide class of one dimensional pair potentials which includes the
 $\delta$-fumction as a singular discontinuous case.  The most general
 pair potential may be derived from an alternative form of (5) namely:

\begin{equation}
(\varphi(x)+\varphi(y)+\varphi(z))^2=g(x)+g(y)+g(z)
\end{equation}
with\hskip3truecm  $x+y+z=0$\hskip1truecm and \hskip1truecm  
$g(x)=\varphi^2(x)+2f(x)$.\\

Equation(7) is in fact a property of the Weierstrassian elliptic functions $\zeta(x)$
 and $P(x) ^{10)}$
\begin{equation}
(\zeta(x)+\zeta(y)+\zeta(z))^2=P(x)+P(y)+P(z)
\end{equation}
with \hskip4truecm $x+y+z=0$ \\

As pointed out by Sutherland$^{8)}$ the effective pair potential 
\begin{equation}
V_{2}^{*}(x)=\lambda[\zeta'(x)+\lambda\zeta^2(x)]-\lambda^2P(x)
\end{equation}
contains all the known soluble pair potentials as limiting cases $^{3,7)}$.

The philosophy of the Sutherland's condition is similar to the one\\ 
advocated in the theory of integrable quantized fields in 1+1 dimensions
 in which the N-body S-matrix is assumed to be factorizable into
 a product of 2-body S-matrices. In fact there exists a relation between
 the two classes of theories since the non relativistic limit of a 
particular S-matrix theory is precisely the phase shift in a pair
 potential of one the type $V_{2}^{*}$ as shown by Zamolodchikov and 
Zamolodchikov $^{9)}$.\\

The corresponding Jastrow wave function turns out to be a product of
 the odd-theta function $\Theta_1$:$^{10)}$
\begin{equation}
\Psi= c\prod_{\scriptstyle i<j}\Theta_1(\displaystyle\frac{\pi}{L}(x_i-x_j))
\end{equation}
which has served to demonstrate that a quantum plasma in one dimension
exhibits long range crystalline order $^{8)}$.
 We shall come back to this wave function later on.\\

3- It is therefore of interest to raise the question whether in two
 dimensions a Jastrow wave function would exhibit also such nice features.

For convenience we use complex coordinates in the plane: $z=x+iy$,\\
$ \overline{z}=x-iy $ and consider the pair wave function $\psi(z,
 \overline{z})$ as well as 
its partial logarithmic derivatives:
\begin{equation}
\begin{array}{lr}
\vspace{.5truecm}
\varphi(z, \overline{z})=\displaystyle\frac{\partial}{\partial z}\log
\psi(z, \overline{z}) & \\
\vspace{.5truecm}
\overline{\varphi}(z, \overline{z})=\displaystyle\frac{\partial}
{\partial \overline{z}}\log\psi(z, \overline{z}) & \\
\end{array}
\end{equation}
 
Following the same line of argumentation we find that the Jastrow wave
 function:
\begin{equation}
\Psi=\displaystyle\prod_{\scriptstyle i<j}|\psi(z_i-z_j,\overline{z}_i
-\overline{z}_j)|^\lambda
\end{equation} 
describes an assembly of N particles experiencing only effective pair
 potentials of a  form analogous to (6):i.e 
\begin{equation}
V_{2}^{*}(z,\overline{z})=\lambda\{\displaystyle\frac{\partial^2}{\partial z
\partial \overline{z}}\log \psi+\lambda\varphi(z, \overline{z})\overline
{\varphi}(z, \overline{z})-\lambda f(z, \overline{z})\}
\end{equation}

if a two-dimensional Sutherland's condition is satisfied:
\begin{equation}
\begin{array}{lr}
\hskip1truecm \{\hskip.35truecm \varphi(x,\overline{x})
\overline{\varphi}(y,\overline{y}) \hskip0.5truecm +\hskip0.5truecm
  \varphi(y,\overline{y})\overline{\varphi}(x,\overline{x}) & \\
\hskip1truecm +\hskip0.3truecm \varphi(y,\overline{y})\overline{\varphi}(z,\overline{z})\hskip0.5truecm +\hskip0.5truecm
  \varphi(z,\overline{z})\overline{\varphi}(y,\overline{y}) & \\
\hskip1truecm +\hskip0.3truecm \varphi(z,\overline{z})
\overline{\varphi}(x,\overline{x})\hskip0.5truecm +\hskip0.5truecm 
  \varphi(x,\overline{x})\overline{\varphi}(z.\overline{z})
\hskip.35truecm \} & \\ 
\hskip4truecm = [ f(x,\overline{x}) + f(y,\overline{y}) +
 f(z,\overline{z}) ]& \\
\end{array}
\end{equation}
\hskip1truecm with:$$ x+y+z=0\hskip1truecm  {\rm and}
\hskip1truecm \overline{x}+\overline{y}+\overline{z}=0 $$

However the analog of eq.(7) in two dimensions:
\begin{equation}
\begin{array}{lr}
\{\varphi(x,\overline{x})+\varphi(y,\overline{y})+\varphi(z,
\overline{z})\}\{\overline{\varphi}(x, \overline{x})+\overline{\varphi}
(y,\overline{y})+\overline{\varphi}(z,\overline{z})\} & \\
\hskip4truecm  =g(x,\overline{x})+g(y,\overline{y})+g(z,\overline{z}) & \\
\end{array}
\end{equation}
together with the constraints on the arguments is not known to be satisfied by two variable
 generalizations
of the weierstrassian elliptic $\zeta_j(z, \overline{z})$ and $P_{ij}(z,
 \overline{z})$ functions$^{ 11)}$. Let us observe that by
 considering a toy model for which the grownd state is represented
 by the Jastrow function:

\begin {equation}
\Psi=c\prod_{\scriptstyle i<j}\Theta_1(\frac{\pi}{L}(x_i-x_j))
\Theta_1(\frac{\pi}{L'}(y_i-y_j))
\end{equation}
as a two dimensional version of (10) where $L$ and $L'$ are lengths in the $x$
and $y$ directions. Then we may verify that
\begin{eqnarray}
\varphi(z,\overline{z})=\frac{1}{2}(\zeta(x)-\zeta(w)\frac{x}{w})-\frac{i}{2}
(\zeta(y)-\zeta(w')\frac{y}{w'}) & \\
\overline{\varphi}(z,\overline{z}) 
=\frac{1}{2}
(\zeta(x)-\zeta(w)\frac{x}{w})+\frac{i}{2}
(\zeta(y)-\zeta(w')\frac{y}{w'}) & 
\end{eqnarray}

\noindent
where $w$,$ w'$ are the periods of $P(x)$, fulfill (15) in an obvious
nice way:


\begin{equation}
\begin{array}{lr}
\hskip.4truecm\frac{1}{2}\{[\zeta(x_1)+\zeta(x_2)+\zeta(x_3)]-i[\zeta(y_1)+\zeta(y_2)
+\zeta(y_3)]\} & \\

\hskip.4truecm\frac{1}{2}\{[\zeta(x_1)+\zeta(x_2)+\zeta(x_3)]+i[\zeta(y_1)+\zeta(y_2)
+\zeta(y_3)]\} & \\

 =\frac{1}{4}\{ [\zeta(x_1)+\zeta(x_2)+\zeta(x_3)]^2+[\zeta(y_1)+
\zeta(y_2)+\zeta(y_3)]^2\} & \\

=\frac{1}{4}\{[P(x_1)+P(x_2)+P(x_3)]+[P(y_1)+P(y_2)+P(y_3)]\}
\end{array}
\end{equation}
since $x_1+x_2+x_3=y_1+y_2+y_3=0 $ and because of (8).

In fact this bidimensional toy model is made up of two independent
 Sutherland's model in perpendicular directions. It is rather unphysical
since the particles are confined in ``quadrants'' relative to each other
and are forced them to keep a constant relative order in the plane.\\

Let us now examine some properties of eqs(14,15).\\

a) The two dimensional Sutherland's condition remains globally invariant under
 the double substitution:
\begin{equation}
\begin{array}{lr}
\varphi(x,\overline{x})\to\varphi(x,\overline{x})+ax+a'\overline{x}+b & \\
\overline{\varphi}(x,\overline{x})\to\overline{\varphi}(x,\overline{x})
+\overline{a'}x+\overline{a}\overline{x}+\overline{b} 
\end{array}
\end{equation}

This transformation induces a change in the effective pair potential\\
 according to:
\begin{equation}
\begin{array}{lr}
V^*\to\{(V^*+a')+\varphi(x,\overline{x})(a'x+\overline{ax}+\overline{b}) & \\
\hskip3.2truecm +\overline{\varphi}(x,\overline{x})(a'\overline{x}+ax+b)\}. & \\
\end{array} 
\end{equation}

One may thus generates new potentials $V^{*}$ from known ones.\\

b) If $\psi$ is such that $\displaystyle\frac{\partial^2}{\partial z\partial\overline{z}}\log\psi=0$, then $\log\psi=f(z)+g(\overline{z})$,where $ f $ and $g$ are independent functions not related to those of eq.(5) or (7). Consequently the Sutherland's condition becomes simpler:
\begin{equation}
\{f'(x)+f'(y)+f'(z)\}\{g'(\overline{x})+g'(\overline{y})+g'
(\overline{z})\}=h(x,\overline{x})+h(y,\overline{y})+h(z,\overline{z}),
\end{equation}
for $x+y+z=\overline{x}+\overline{y}+\overline{z}=0$ and $h$
is a new function.
We observe that if $g'= $const ( or resp. $f'=$const ) the Sutherland's
 condition  is automatically satisfied . Moreover if $g'=0$ 
(or resp. $f'=0$) there is no effective pair potential present.
  This is relevant for wavefunctions 
 in the quantum Hall effect as we shall see. \\

c) Let us now consider the physically attractive situation where $\psi$ is
solely function of the distance :$\psi=\psi(z\overline{z})$. If one
 denotes its logarithmic  derivative by $\varphi=\varphi(z\overline{z})$
then the Sutherland's condition becomes:
\begin{equation} 
\begin{array}{lr}
\varphi(y\overline{y})\varphi(x\overline{x})(\overline{x}y+\overline{y}x)+
\varphi(y\overline{y})\varphi(z\overline{z})(y\overline{z}+\overline{y}z) 
+ \varphi(z\overline{z})\varphi(x\overline{x})
(x\overline{z}+z\overline{x}) & \\
\hskip5truecm  = f(x,\overline{x})+f(y,\overline{y})+f(z,\overline{z}), 
\end{array}
\end{equation} 
Note that if $f$ is function of the distance, one has an effective
 potential only dependent on the distance.\\

To get more insight, we introduce vectors through: $\overline{x}y+\overline{y}x=2\vec{x}.\vec{y}$ etc... This leads  to the suggestive interesting form:
\begin{equation}
[\varphi(x)\vec{x}+\varphi(y)\vec{y}+\varphi(z)
\vec{z}]^{2}=g(x)+g(y)+g(z)
\end{equation}
 with $\vec{x}+\vec{y}+\vec{z}=\vec{0},x=|\vec{x}|,y=|\vec{y}| $  and
 $z=|\vec{z}|.$

Would this Sutherland condition admit a non trivial solution? To answer
 this question let us consider the equilateral triangle configuration
 of the three vectors $\vec{x},\vec{y},\vec{z}$ admitting the origin as
 center of mass, namely \\ $|\vec{x}|=|\vec{y}|=|\vec{z}|=x$, then (25)
 reduces to 
 
\begin{equation}
\varphi^2(x)(\vec{x}+\vec{y}+\vec{z})^{2}=3g(x)
\end{equation}
for all $x$.  But,  because of $\vec{x}+\vec{y}+\vec{z}=0$, this implies
 that $g(x)=0$ automatically as a function of $x$.

Hence ultimately we obtain
\begin{equation}
\varphi(x)\vec{x}+\varphi(y)\vec{y}+\varphi(z)\vec{z}=\vec{0}
\end{equation}
as well as $\vec{x}+\vec{y}+\vec{z}=0$. But it is clear that the two
 conditions are in-\\ consistent with each other unless $\varphi(x)$
 is a {\it constant }. This later
case cor-\\responds to the physically relevant displaced harmonic potential
 which has been examined and solved by many authors$^{12)}$. Thus it
 seems very
 much unlikely that the Jastrow ansatz with functions depending only
 on the dis-\\tance can be of use in two dimensions. The two alternative
 ways out are the
 occurence of unphysical pair force which would depend on directions
such as the case in our toy model, or the need of a more general
 ansatz which would take into account three body forces, see for example
 Calogero and Marchioro $^{13)}$.\\

4- In recent years, the Jastrow wave function has become very popular
 because it describes, in the so-called Laughlin wave function for the
 Quantum Hall Effect the relative motion of N planar electrons in their
 lowest Landau level.\\

As shown by Haldane and Rezayi$^{14)}$,  the Jastrow part of the Laughlin
 wave function, in the presence of toroidal boundary conditions takes 
precisely the form of eq. (10), where the abscissas $x_j$ are now
 continued analytically in the complex plane $x_j\to z_j$. However 
 the complex 
conjugate coordinates $\overline{z}_j$ are not present. This  means that 
according to property b (see eq.(23)) the two-dimensional Sutherland's 
condition (15) is trivially fulfilled. And as shown in ref. (15) the
 analyticity in the $z_j$ does not affect the total energy but only the 
angular momentum (and possibly the statistics of the particles)
 carried by the wave function.\\

This connection with the most general (elliptic) one dimensional
 Suther-\\land model$^{8)}$ is not so surprising  since some earlier
 connections with the simpler Calogero-Sutherland model have been
 pointed out $^{16,17)}$. Recent progress on this class of quantum
 integrable systems may, through this connection, shed some new light 
on the Quantum Hall Effect.\\

5- In conclusion, we have attempted to determine  whether or not the
 Jastrow wave function in two dimensions may be consistent with the sole
 existence of two body forces in an  assembly of N particles. This has
 led to a bidimensional generalisation of the Sutherland's condition$^{
7)}$ for the wave function of a pair of particles. If the interparticle 
forces are distance dependent then only attractive harmonic forces
 are physically allowed. Finally we point out the remarkable connection 
 between the Laughlin wave function of the Quantum Hall Effect with 
toroidal boundary condition and the one-dimensional quantum elliptic 
Sutherland's model$^{8)}$.

\end{document}